\documentclass{PoS}

\usepackage{amsmath,amssymb,graphicx,multirow,tabularx,physics}
\usepackage{davitex}

\title{Polyakov Loop Susceptibility and Correlators in the Chiral Limit}

\ShortTitle{Polyakov Loops in the Chiral Limit}

\author{\speaker{David A. Clarke}\\
        Fakult\"at f\"ur Physik, Universit\"at Bielefeld, D-33615
        Bielefeld, Germany\\ 
        E-mail: \email{dclarke@physik.uni-bielefeld.de}}

\author{Olaf Kaczmarek\\
        Fakult\"at f\"ur Physik, Universit\"at Bielefeld, D-33615
        Bielefeld, Germany;\\
        Key Laboratory of Quark \& Lepton Physics (MOE) and Institute of
        Particle Physics, \\Central China Normal University, Wuhan 430079, 
        China\\
        E-mail: \email{okacz@physik.uni-bielefeld.de}}

\author{Frithjof Karsch\\
        Fakult\"at f\"ur Physik, Universit\"at Bielefeld, D-33615
        Bielefeld, Germany\\
        E-mail: \email{karsch@physik.uni-bielefeld.de}}

\author{Anirban Lahiri\\
        Fakult\"at f\"ur Physik, Universit\"at Bielefeld, D-33615
        Bielefeld, Germany\\
        E-mail: \email{alahiri@physik.uni-bielefeld.de}}

\abstract{
In quenched QCD the Polyakov loop is an order parameter of the
deconfinement transition, but with decreasing quark mass, the peak in the
Polyakov loop susceptibility becomes less pronounced, and it loses its
interpretation as an indicator for deconfinement. 
For this $N_f=2+1$ HISQ study, we fix the strange quark mass $m_s$ at its 
physical value and investigate the dependence of the Polyakov loop on 
the light quark mass $m_l$ in the range
$m_s/m_l=27-160$, following $m_l$ toward the chiral limit.
In particular we will look how the inflection point and susceptibility
behave as we decrease $m_l$, to see whether one finds any indication of
a crossover, and therefore whether the Polyakov loop is sensitive to 
the chiral phase transition.
Preliminary results show no signal of a 
crossover from the real part of the Polyakov loop in the vicinity of 
the chiral crossover.
Closely related is an investigation of Polyakov loop correlations and the
Debye mass in this limit. Preliminary results suggest little or no 
dependence on $m_l$.
}

\FullConference{37th International Symposium on Lattice Field Theory - 
                Lattice2019\\
		16-22 June 2019\\
		Wuhan, China}

\begin{document}

\section{The Polyakov loop}

For lattice QCD in a finite volume $N_\sigma^3\times N_\tau$, the thermal 
Wilson line $L_{\vec{x}}$, the Polyakov loop $P_{\vec{x}}$, and its spatial
average $P$ are given by
\begin{equation}\label{eq:ploop}
  L_{\vec{x}}\equiv\prod_\tau U_4(\vec{x},\tau),~~~~~~
  P_{\vec{x}}\equiv\frac{1}{3}\tr L_{\vec x},
  ~~~~\text{and}~~~~
  P\equiv\frac{1}{N_\sigma^3}\sum_{\vec{x}}P_{\vec{x}},
\end{equation}
respectively. 
In general we define a susceptibility $\chi_O$ for an intensive 
observable $O$ as
\begin{equation}
  \chi_O=N_\sigma^3\left(\ev{O^2}-\ev{O}^2\right).
\end{equation}
The Polyakov loop susceptibility and the susceptibility of the real
part of the Polyakov loop are then $\chi_{\,|P|}$ and $\chi_{\,\Re P}$.
At low temperatures with static quarks in the infinite
volume limit, the Polyakov loop expectation value $\ev{|P|}$ is zero due
to the global $\mathbb{Z}_3$ symmetry of the gauge action. At higher
temperatures this center symmetry is spontaneously broken, and $\ev{|P|}$
acquires a nonzero value, signalling a finite static quark-antiquark free
energy at large separations, and hence deconfinement. At finite 
$N_\sigma$, $\ev{|P|}$ as a
function of temperature has an inflection point, and the slope at this
point diverges in the infinite volume limit. The susceptibility as a
function of temperature, meanwhile, exhibits a pronounced peak 
at finite $N_\sigma$ whose height diverges in the infinite volume 
limit as $\chi_{\,|P|}^{\rm max}\sim N_\sigma^3$,
reflecting the first order nature of the deconfinement phase transition
in pure $\SU(3)$ gauge theory.

At finite quark mass, the Polyakov loop is no longer a strict order
parameter; the finite quark mass breaks the $\mathbb{Z}_3$ symmetry
explicitly, and $\ev{|P|}$ and $\ev{\Re P}$ are never zero. 
Nevertheless at larger-than-physical
quark mass, some remnants of critical behavior seem to remain, and in
particular, past studies have found inflection points for $\ev{|P|}$ and the
chiral condensate to appear at similar temperatures. For example a 2008
study~\cite{cheng_qcd_2008} using improved staggered fermions of
$N_f=2+1$ flavors, with light quark masses corresponding
to about a 50\% larger-than-physical pion mass, performed on coarse lattices
with $N_\tau=4$ and $6$, found inflection points in both $\ev{|P|}$
and the chiral order parameter to lie around the same temperature. 

\begin{figure}[t]
\centering
\includegraphics[width=0.88\textwidth]{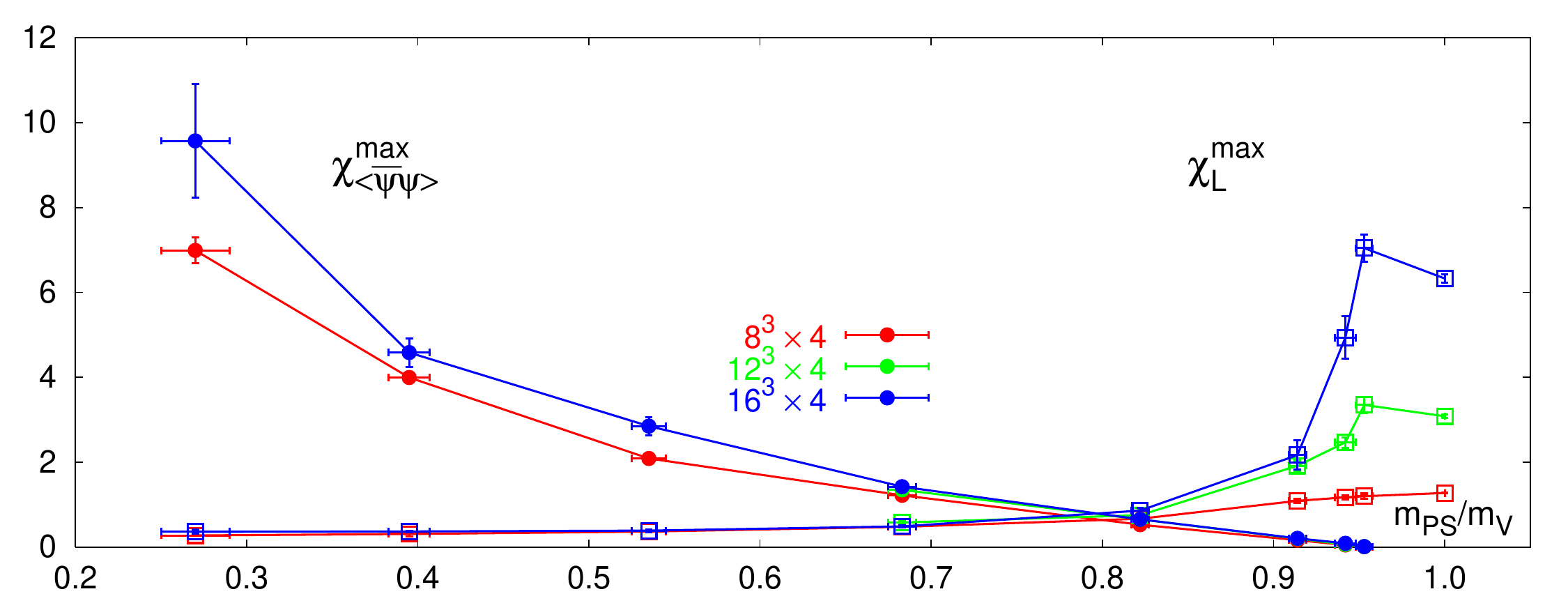}
\caption{Dependence of peak heights of Polyakov loop and chiral
         susceptibilities on $N_\tau=4$ lattices with $N_\sigma=8$, 12, and 16
         as a function of the pseudo-scalar mass in units of the vector meson
         mass, which is another way of stating quark mass. This calculation
         used a standard staggered fermion discretization scheme. Figure taken
         from Ref.~\cite{plessas_lattice_2002}.}
\label{fig:dec_suscept_peak}
\end{figure}

While these earlier studies of Polyakov loop expectation values
seemed to clearly indicate that the inflection point of the Polyakov loop
is related to the chiral transition, only the latter has a clear-cut
interpretation as a phase transition in the chiral limit.
This ``coincidence'' of inflection points in the temperature dependence of
$\ev{|P|}$ and the chiral condensate is often taken as evidence for the
coincidence of a chiral and deconfinement transition.
To what extent deconfinement or the melting of bound states in general can
be associated with properties of the Polyakov loop at finite quark mass
is, however, an open question. 
In fact, studies with improved fermion actions in general showed
that the QCD transition, which is a pseudo-critical crossover
transition at non-zero values of the quark masses, tends to
become a {\it smoother} transition when going closer to the
continuum limit and using more highly improved fermion actions
such as HISQ or stout actions~\cite{bazavov_chiral_2012}.
Indeed a potential hint that this
behavior weakens was already given in Ref.~\cite{plessas_lattice_2002},
which we show in Figure~\ref{fig:dec_suscept_peak},
where ones sees that $\chi_{\,|P|}^{\rm max}$ decreases with 
decreasing quark mass
for fixed $N_\tau$. 
Moreover there are studies with more highly improved actions
that do not find the inflection point
of $\ev{|P|}$ to coincide with the chiral inflection 
point~\cite{aoki_qcd_2009,bazavov_polyakov_2016}, 
challenging the notion of using the Polyakov loop as an observable for 
deconfinement of light degrees of freedom.
In actuality, then, it is not clear to what
extent properties of the Polyakov loop can provide a reasonable
criterion for deconfinement. 

Therefore one aim of the present study is to investigate whether an
indication of deconfinement from the Polyakov loop, if any, weakens 
as we lower the light quark mass $m_l$. We employ an improved
staggered action (HISQ) that leads to greatly reduced taste violations,
thus providing a better approach to the continuum limit. 
We analyze the
Polyakov loop and Polyakov loop correlation functions on lattices with
temporal extent $N_\tau =8$ and 12, which in other thermodynamics
calculations with HISQ have been shown to provide results close to the
continuum limit~\cite{bazavov_equation_2014}.

\section{Debye screening}
The Polyakov loop relates to $F_{q\bar{q}}$,
the color-averaged free energy of a static quark-antiquark pair in
equilibrium at temperature $T$, by
\begin{equation}\label{eq:Fav}
  \exp\left[-F_{q\bar{q}}(r,T)/T\right]
    =\ev{P_{\vec{x}}\,P^\dagger_{\vec{y}}}
   \stackrel{r\to\infty}{\sim}|\ev{P}|^2=\ev{\Re P}^2,
     \qquad rT=\left|\vec{x}-\vec{y}\tinysp\right|/a\,N_\tau,
\end{equation}
because the expectation value of the Polyakov loop lies in the real 
$\mathbb{Z}_3$ sector for finite quark masses.
The gauge-invariant color-averaged Polyakov loop correlator can be decomposed
into (in general gauge-dependent) color singlet and color octet contributions
\cite{mclerran_monte_1981,mclerran_quark_1981,nadkarni_non-abelian_1986}.
In particular the color singlet
\begin{equation}
  F_1(r,T)=-T\log\ev{\frac{1}{3}\tr L_{\vec{x}}\,L_{\vec{y}}^{\dagger}},
\end{equation}
which clearly depends on the gauge. Hence it is important to restrict to
a particular gauge before measuring $F_1$.
In the deconfined phase, the interaction between two charges is screened
by the medium. The distance at which in-medium modifications of the
quark-antiquark interaction dominate is characterized by the Debye screening
radius $r_D$. Its inverse, the Debye screening mass $m_D$,
can be extracted from the long-distance ($rT\gg 1$) behavior of $F_1$ as
\begin{equation}
  F_1(r,T)\simeq-\frac{4}{3}\frac{\alpha(T)}{r}e^{-r\,m_D(T)}+F_1(r=\infty,T).
\end{equation}
The Debye mass is known to depend on $N_f$ at high temperatures through
a leading order perturbative calculation. At lower temperatures, lattice
calculations still show qualitative agreement with perturbation
theory~\cite{kaczmarek_screening_2008}. The Debye mass will also depend on
$m_l$; hence another goal of this study will be to see how
$m_D$ changes with $m_l$.

\section{Setup and simulation parameters}

\begin{table}
\begin{tabularx}{\linewidth}{LCC|CCR} \hline\hline
$N_\sigma^3\times N_\tau$ & $m_s/m_l$ & avg. \# TU &
$N_\sigma^3\times N_\tau$ & $m_s/m_l$ & avg. \# TU \\
\hline
$32^3\times8$  & 27  & 62 000 & $60^3\times12$ & 40  & 30 000\\
$40^3\times8$  & 40  & 52 000 &                & 80  & 17 000\\
$56^3\times8$  & 80  & 20 000 &                &     &       \\
               & 160 & 17 000 &                &     &       \\
\hline\hline
\end{tabularx}
\caption{Summary of parameters used in these proceedings and corresponding
         statistics, reported in average molecular dynamic time units (TU)
         per parameter combination.}
\label{tab:latsnstats}
\end{table}
We use for our analysis configurations that were generated with the
HISQ action in $(2+1)$-flavor QCD with a physical value of the strange quark
mass and light quark masses in the range $m_l=m_s/27$ to $m_s/160$,
corresponding to $140~\text{MeV}\gtrsim m_\pi\gtrsim 58~\text{MeV}$
\cite{ding_chiral_2019}.
The bare coupling $\beta$ is taken in the range 6.260-6.850 depending
on the lattice, which was chosen so that the temperatures would lie
in the vicinity of the chiral pseudo-critical temperature.
Measurements of the free energies are made in the Coulomb gauge, with
the gauge fixing carried out using the over-relaxation
algorithm~\cite{mandula_efficient_1990}.
The scale has been set using the experimental value
$f_K=156.1/\sqrt{2}$~MeV~\cite{beringer_review_2012} as well as
the updated results for the kaon decay constant calculated with the
HISQ action, $f_K\,a(\beta)$, given in Ref.~\cite{bazavov_meson_2019}.
We renormalize the Polyakov loop using the 
$qq$-scheme~\cite{kaczmarek_heavy_2002}, where the $T=0$
potential
is obtained from a fit of the combined data shown in Fig.~14
of Ref.~\cite{bazavov_equation_2014} using the ansatz
\begin{equation}
  V_{T=0}(r)=\frac{A}{r}+B+Cr.
\end{equation}
Additive renormalization constants have been calculated at $(r/a)^2=4$. 
Distances up to $(r/a)^2=13$
have been tree-level corrected~\cite{necco_nf0_2002}.
Error bars are calculated in 32 jackknife bins unless otherwise stated.
The lattice sizes, quark masses, and statistics are summarized in
Table~\ref{tab:latsnstats}.

\section{Results}
\begin{figure}
\centering
\includegraphics[width=0.48\textwidth]{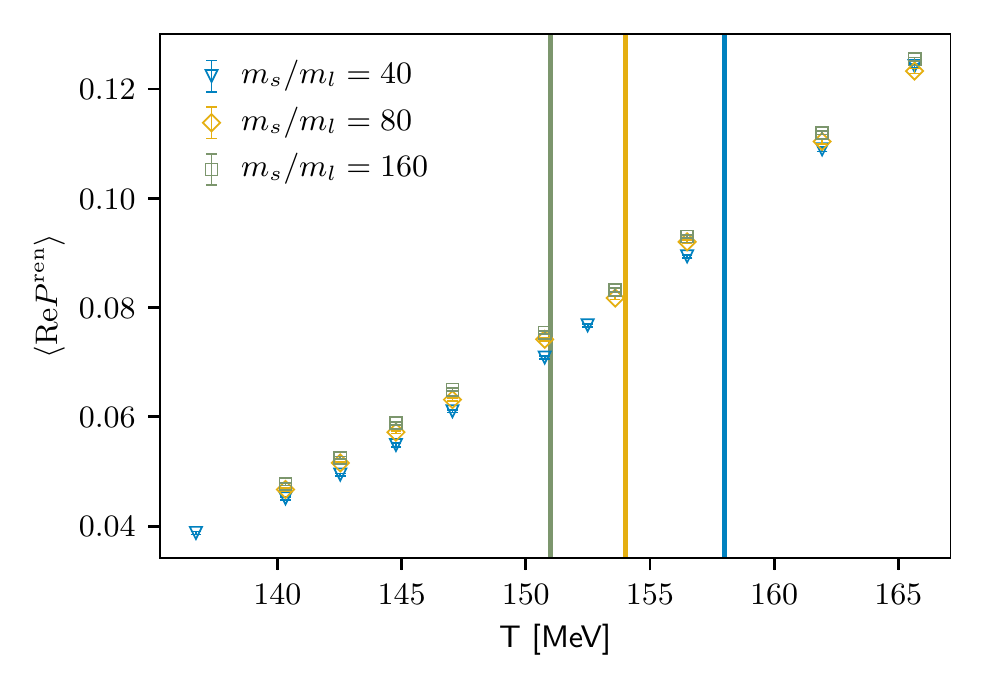}
\includegraphics[width=0.48\textwidth]{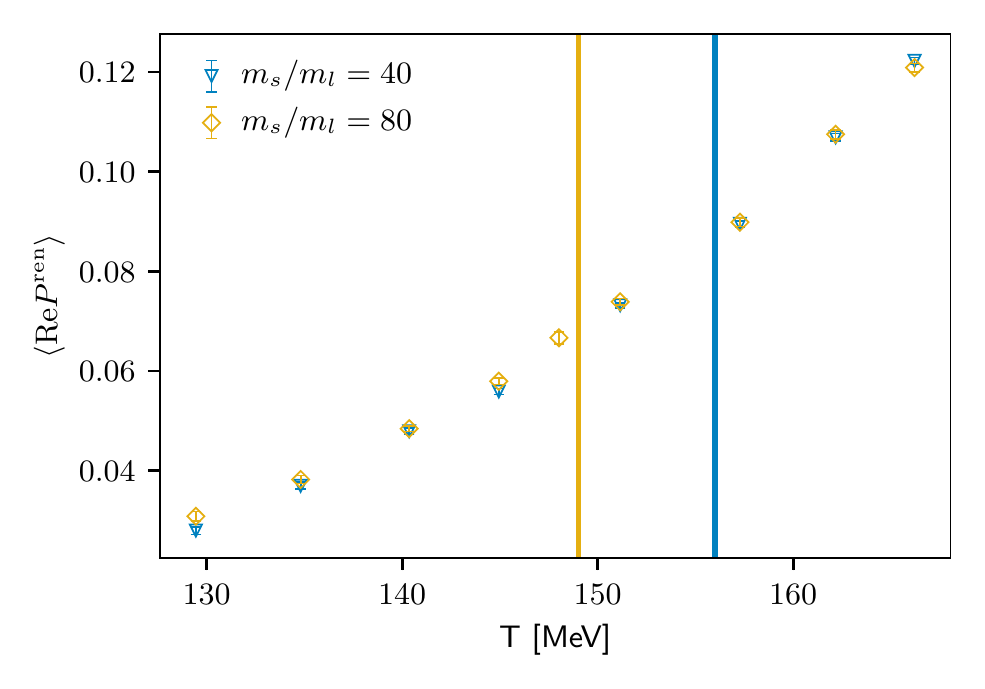}
\includegraphics[width=0.48\textwidth]{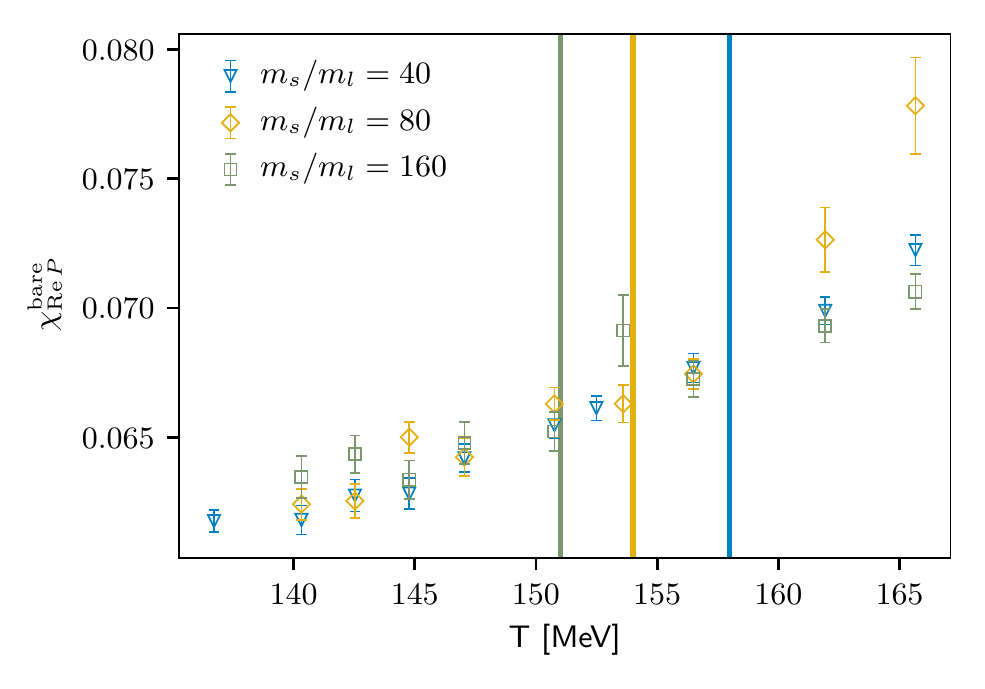}
\includegraphics[width=0.48\textwidth]{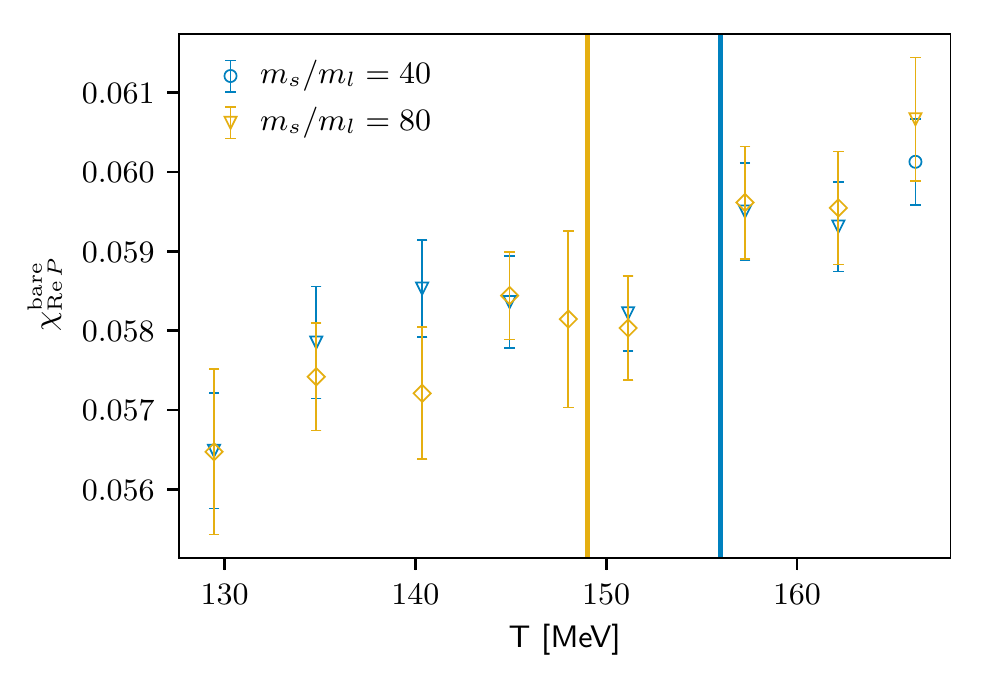}
\caption{Real part of the renormalized Polyakov loop (top) and the 
         susceptibility of the real part of the bare Polyakov loop (bottom) 
         near the chiral pseudo-critical temperature for $N_\tau=8$ (left) and 
         $N_\tau=12$ (right) for different $m_s/m_l$. The vertical lines 
         indicate the chiral pseudo-critical temperatures corresponding to 
         each quark mass ratio.}
\label{fig:susc_reno}
\end{figure}

The real part of the renormalized Polyakov loop is shown for different 
values of $m_s/m_l$
in the two plots in the top row of Figure~\ref{fig:susc_reno} 
for $N_\tau=8$ (left) and $N_\tau=12$ (right) lattices. 
We plot this instead of $\ev{|P^{\text{ren}}|}$ because 
$\ev{|P^{\text{ren}}|}$ receives significant finite volume corrections 
from $\Im P$, and because after explicit breaking of the
$\mathbb{Z}_3$ symmetry by finite quark mass, the system settles in a region 
where $P$ is centered on the real axis. 
Vertical lines indicate the positions of chiral
pseudo-critical temperatures taken from Ref.~\cite{ding_chiral_2019}, which
are at roughly 151, 154, and 158~MeV for $N_\tau=8$, and 149 and
156~MeV for $N_\tau=12$. 
In the entire temperature range probed by us
$\ev{\Re P^{\text{ren}}}$ is convex for both $N_\tau$. 
Calculations performed with other lattice sizes similarly show no
indication that a crossover signal from the Polyakov loop coincides with
the signal from the chiral condensate.
An inflection point, which needs to exist as $\ev{|P^{\text{ren}}|}$ 
(or equivalently at $N_\sigma=\infty$, $\ev{\Re P}$) will 
eventually approach unity at high temperature, will therefore only occur at 
temperatures larger than $1.1\,T_{pc}$.
For $N_\tau=8$, decreasing $m_l$ leads to an
increase in $\ev{\Re P^{\text{ren}}}$ across all probed temperatures, with
no indication that the slope changes.
Meanwhile for $N_\tau=12$ there is no $m_l$ dependence within the statistics.

The two plots in the bottom row of Figure~\ref{fig:susc_reno} show
$\chi^{\text{bare}}_{\,\Re P}$, the susceptibility of the real part
of the bare Polyakov loop, for different $m_l$ for $N_\tau=8$ (left) 
and $N_\tau=12$ (right) for the same configurations. 
We see no peak in this range, which is consistent with what we find 
for $\ev{\Re P^{\text{ren}}}$. There is no sensitivity to $m_l$ for either 
$N_\tau$ in this range of temperatures.
Our preliminary results thus show no overlap of a signal from the Polyakov loop
with the chiral condensate. Higher
statistics and more results at higher temperatures
are needed to clearly see the effect of $m_l$ on $\chi_{\,\Re P}^{\text{max}}$.


\begin{figure}[t]
\centering
\includegraphics[width=0.48\textwidth]{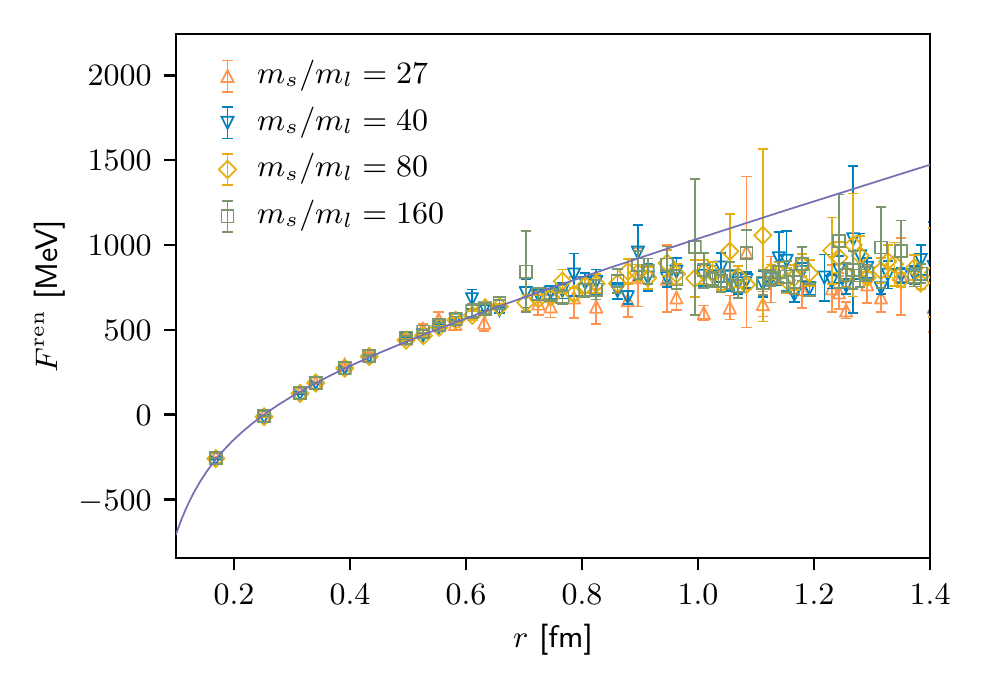}
\includegraphics[width=0.48\textwidth]{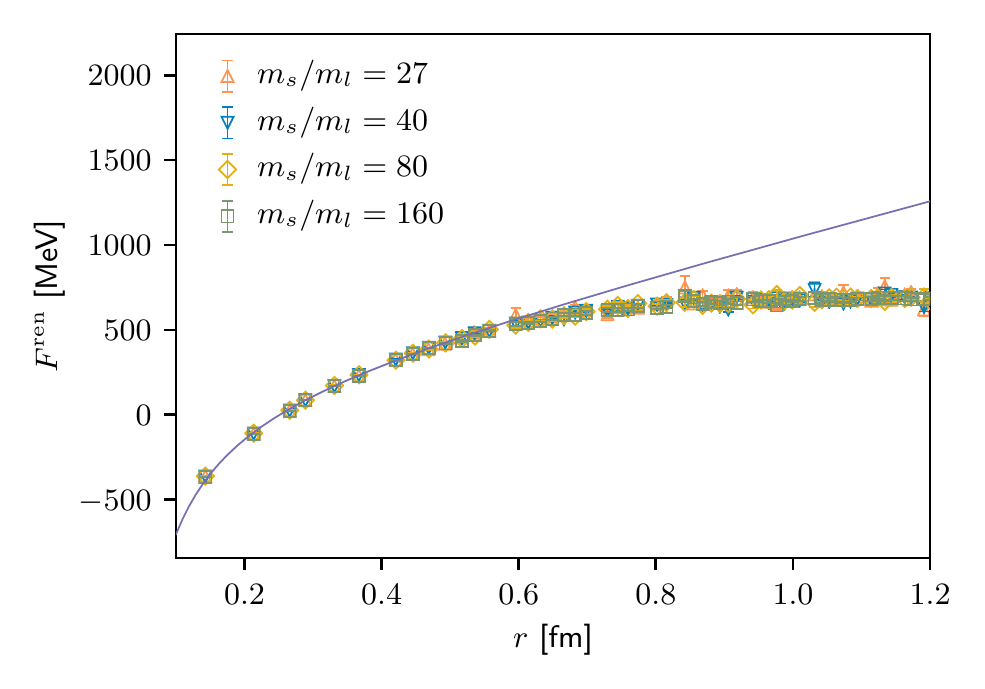}
\caption{Dependence of $F_1$ on $m_l$ for $N_\tau=8$ and our largest
         $N_\sigma$ for $\beta=6.285$ or $T=141$ [MeV] (left) and 
         $\beta=6.445$ or $T=166$ [MeV] (right).
         The purple line indicates the zero temperature potential.}
\label{fig:mdepend}
\end{figure}

In Figure~\ref{fig:mdepend} we plot the renormalized singlet free energy against
the Polyakov loop separation $r$ in physical units at $N_\tau=8$ for different
$m_s/m_l$. The left plot shows results at $\beta=6.285$, which corresponds
to a temperature of about 141~MeV, while the right plot has $\beta=6.445$, which
is at about 166~MeV. The data are given at our largest available $N_\sigma$
for each $\beta$ and $m_l$ combination, which we find are large enough
to suppress finite volume effects. For both plots one can see by eye that the
long-distance results for all $m_l$ agree within statistical uncertainty.
This suggests that the Debye screening masses, eventually extracted from
these correlation functions, show little or no dependence on the light
quark masses.

Extraction of $m_D$ has not yet yielded any precise results, as $F_1$ tends
to be rather noisy away from short distances, which has made carrying out
a long-distance fit difficult. We plan to try smoothing our configurations
using the gradient flow~\cite{luscher_properties_2010},
which mostly affects
only short-distance physics, and could therefore improve our long-distance
signal without spoiling it.

\section{Summary and Outlook}
We presented here some first results for our research on indicators of
hadron melting toward the chiral limit. The Polyakov loop does not exhibit
any indication of a crossover near the chiral pseudo-critical temperature
at lower-than-physical quark mass, consistent with the results of more
recent HISQ studies at physical $m_l$. 
More results at other $m_l$ are forthcoming, which will allow us to examine
the behavior of $\chi_{\,\Re P}^{\text{max}}$ and the slope of $\ev{\Re P}$ in
the chiral limit. Further results at other $N_\tau$ will also be analyzed,
allowing for a continuum limit extrapolation.
We are working toward a determination of $m_D$,
however statistical noise in the free energy at large distances make the
calculation difficult. 
Our current data suggest $m_D$ will have no or little dependence on $m_l$
in the range investigated. 
Smoothing using the gradient flow may be attempted in the future.

\acknowledgments
This work was supported by the Deutsche Forschungsgemeinschaft (DFG,
German Research Foundation) - project number 315477589 - TRR 211; and from
the German Bundesministerium f\"ur Bildung und Forschung through
Grant No. 05P18PBCA1.
We thank HotQCD for providing access to their latest data
sets and for many fruitful discussions.
\pagebreak

\bibliographystyle{JHEP}
\bibliography{bibliography}

\end{document}